\documentclass[aps,prl,twocolumn,amsmath,amssymb,superscriptaddress]{revtex4-2}
\usepackage{graphicx} 
\usepackage{xcolor}
\usepackage{threeparttable}
\usepackage[colorlinks=true,
					linkcolor=blue,
					anchorcolor=blue,
					citecolor=blue,
					urlcolor=magenta]{hyperref}
\usepackage{hyperref}
\usepackage[normalem]{ulem}

\usepackage{bbold}

\newcommand{\bra}[1]{\ensuremath{\left\langle#1\right|}}
\newcommand{\ket}[1]{\ensuremath{\left|#1\right\rangle}}

\begin{document}


\title{Dressed-state relaxation in coupled qubits as a source of two-qubit gate errors}

\newcommand{\BAQIS}{\affiliation{1}{Beijing Key Laboratory of Fault-Tolerant Quantum Computing, Beijing Academy of Quantum Information Sciences, Beijing 100193, China}}

\newcommand{\IOP}{\affiliation{2}{Beijing National Laboratory for Condensed Matter Physics and Institute of Physics, Chinese Academy of Sciences, Beijing 100190, China}}

\newcommand{\UCAS}{\affiliation{3}{University of Chinese Academy of Sciences, Beijing 101408, China}}

\newcommand{\HFNL}{\affiliation{4}{Hefei National Laboratory, Hefei 230088, China}}

\author{Ruixia Wang}
\thanks{These authors have contributed equally to this work.}
\affiliation{\BAQIS}

\author{Jiayu Ding}
\thanks{These authors have contributed equally to this work.}
\affiliation{\BAQIS}

\author{Chenlu Wang}
\thanks{These authors have contributed equally to this work.}
\affiliation{\BAQIS}

\author{Yujia Zhang}
\affiliation{\BAQIS}
\affiliation{\IOP}
\affiliation{\UCAS}

\author{He Wang}
\affiliation{\BAQIS}

\author{Wuerkaixi Nuerbolati}
\affiliation{\BAQIS}

\author{Zhen Yang}
\affiliation{\BAQIS}

\author{Xuehui Liang}
\affiliation{\BAQIS}

\author{Weijie Sun}
\affiliation{\BAQIS}

\author{Haifeng Yu}
\affiliation{\BAQIS}
\affiliation{\HFNL}

\author{Fei Yan}
\email{yanfei@baqis.ac.cn}
\affiliation{\BAQIS}


\begin{abstract}

Understanding error mechanisms in two-qubit gate operations is essential for building high-fidelity quantum processors. While prior studies predominantly treat dephasing noise as either Markovian or predominantly low-frequency, realistic qubit environments exhibit structured, frequency-dependent spectra. Here we demonstrate that noise at frequencies matching the dressed-state energy splitting—set by the inter-qubit coupling strength $g$—induces a distinct relaxation channel that degrades gate performance. Through combined theoretical analysis and experimental verification on superconducting qubits with engineered noise spectra, we show that two-qubit gate errors scale predictably with the noise power spectral density at frequency $2g$, extending the concept of $T_{1\rho}$ relaxation to interacting systems. This frequency-selective relaxation mechanism, universal across platforms, enriches our understanding of decoherence pathways during gate operations. The same mechanism sets coherence limits for dual-rail or singlet–triplet encodings.

\end{abstract}

\maketitle

High-fidelity two-qubit gates remain a central challenge for scalable quantum information processing across platforms \cite{knill_quantum_2005,preskill_quantum_2018,quantum_2019}. Despite increasingly sophisticated control, residual dissipation and dephasing set a floor to gate errors and complicate the path to fault tolerance \cite{van_vu_fidelity-dissipation_2024,google_quantum_ai_and_collaborators_2025-quantum_2025}. It is therefore essential to identify the physical noise channels that potentially limit performance under realistic gate conditions \cite{error_2024,omalley_qubit_2015}. Because practical noise environments exhibit structured, frequency-dependent spectra rather than idealized white or quasi-static noise \cite{spinlocking2013,sung_2019-non-gaussian,von_lupke_2020-two-qubit,kakuyanagi_2007-dephasing}, understanding how gate errors depend on the frequency content of environmental fluctuations is particularly important. This enables principled error budgeting, targeted mitigation, and hardware-informed gate design.

\begin{figure}[tbhp]
\centering
\includegraphics[scale=1]{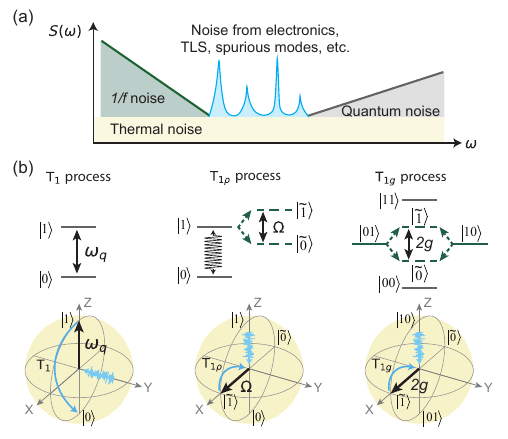}
\caption{\label{fig:scheme}
(a) Schematic diagram of a realistic noise spectrum $S(\omega)$ that includes $1/f$ noise, thermal noise, quantum noise, and spurs from electronics, modes, or microscopic defects. 
(b) Comparison of (left) the energy ($T_1$) relaxation process, (center) the rotating-frame ($T_{1\rho}$) relaxation process, and (right) the two-qubit dressed-state ($T_{1g}$) relaxation processes. 
}
\end{figure}

Common analyses usually assume either Markovian white noise or Johnson-Nyquist noise, which obscures frequency-selective effects during entangling interactions \cite{universal_2022,bialczak_1_2007,yoshihara_2006-decoherence,koch_2007-charge-insensitive}. For example, in superconducting quantum circuits, spurious tones, environmental modes, and microscopic two-level defects~\cite{muller_towards_2019} can rise above the background noise floor in the MHz range (Fig.~\ref{fig:scheme}(a)), which are known to cause relaxation during driven evolution, as shown by spin-locking experiments where the relaxation rate tracks the noise power spectral densities at the Rabi frequency \cite{spinlocking2013}. Yet, in the absence of an external drive, whether---and how---these spectral features impact two-qubit gates has remained unclear.

Here we show that when two coupled qubits are tuned into resonance (as in an iSWAP-type gate) \cite{foxen_2020-demonstrating}, the single-excitation manifold forms a dressed two-level system that is selectively sensitive to the differential-detuning ($\Delta$) or dephasing noise at the dressed splitting $2g$, directly analogous to rotating-frame relaxation in a driven qubit. 
We derive that the dressed-state relaxation rate is given by $\Gamma_g=\tfrac{1}{2}S_{\Delta}(2g)$, where $S_{\Delta}(\omega)$ is the power spectral density (PSD) of the detuning noise. We then map this decoherence channel to two-qubit gate errors for a gate of interaction time $\tau$, obtaining $\epsilon_{\mathrm{avg}}=\tfrac{1}{5}\Gamma_g\tau$ for the average error and $\epsilon=\tfrac{1}{4}\Gamma_g\tau$ for the Pauli error, valid in the weak-noise limit.
Using superconducting qubits with engineered noise spectra, we experimentally validate these relationships by performing dressed-state relaxation and two-qubit gate benchmarking as a function of noise power, $g$, and $\tau$.
Our results confirm the predicted scaling and reveal a gate-activated, frequency-selective decoherence mechanism. These insights apply directly to dual-rail qubit encoding~\cite{shim_2015-semiconductor-inspired,campbell_2020-universal,kubica_erasure_2023,teoh_dual-rail_2023,levine_2024-demonstrating,huang_logical_2025,chou_2024-superconducting} and semiconductor singlet-triplet encoding~\cite{johnson_tripletsinglet_2005,PhysRevLett.110.146804,connors_charge-noise_2022}, where the same dressed-state physics governs error processes, providing guidance for optimizing coupling strengths and noise filtering in quantum circuits.

To build intuition for dressed-state relaxation in a pair of coupled qubits, it is helpful to recall the connection to ordinary energy ($T_1$) relaxation and the rotating-frame ($T_{1\rho}$) relaxation of a driven qubit. As illustrated in Fig.\,\ref{fig:scheme}(b), Fermi’s golden rule (or equivalently Bloch–Redfield theory) states that a qubit’s energy relaxation rate $\Gamma_1=1/T_1$ is set by the \emph{transverse} noise at the qubit transition frequency $\omega_q$, i.e.\ noise components that are perpendicular to the qubit’s quantization axis and resonant with the energy splitting.

\paragraph{$T_{1\rho}$ analogy.}
When a qubit is subject to a resonant drive with Rabi frequency $\Omega$, the drive defines the quantizing field in the rotating frame,
\begin{equation}
\tilde{H} = \frac{1}{2}\Omega\left(\ket{0}\!\bra{1}+\ket{1}\!\bra{0}\right) = \frac{1}{2}\Omega\,\sigma_x \;,
\end{equation}
which can be considered a pseudo-spin (rotating-frame qubit) with energy splitting $\Omega$.
Noise \emph{transverse} to this driving field at frequency $\Omega$ induces transitions between the pseudo-spin eigenstates $\ket{\pm X}=(\ket{0}\pm\ket{1})/\sqrt{2}$, giving the rotating-frame or $T_{1\rho}$ relaxation measurable in the spin-locking experiments \cite{spinlocking2013}. In particular, ordinary dephasing ($z$) noise (longitudinal to the undriven qubit but transverse to the pseudo-spin) contributes to the rotating-frame relaxation rate $\Gamma_{1\rho} = 1/T_{1\rho} = \tfrac{1}{2}\Gamma_1+\Gamma_{\Omega}$ an extra term $\Gamma_{\Omega}=\tfrac{1}{2}S(\Omega)$, where $S(\omega)$ is the power spectral density (PSD) of the dephasing noise. 

\paragraph{Quantization of two coupled qubits.}
Now consider two qubits with bare frequencies $(\omega_1,\omega_2)$ and exchange coupling $g$,
\begin{multline}
H=\frac{1}{2}\omega_1\left(\ket{1}\!\bra{1}-\ket{0}\!\bra{0}\right)\otimes\mathbb{1}
+\frac{1}{2}\omega_2\,\mathbb{1}\otimes\left(\ket{1}\!\bra{1}-\ket{0}\!\bra{0}\right)  \\
+g\left(\ket{01}\!\bra{10}+\ket{10}\!\bra{01}\right).
\end{multline}
Upon moving to the rotating frame and restricting to the single-excitation subspace $\{\ket{01},\ket{10}\}$, the effective Hamiltonian is
\begin{equation}
\tilde{H}=\frac{1}{2}\Delta\!\left(\ket{10}\!\bra{10}-\ket{01}\!\bra{01}\right)
+g\left(\ket{01}\!\bra{10}+\ket{10}\!\bra{01}\right) \;,
\end{equation}
where $\Delta=\omega_1-\omega_2$ is the qubit-qubit detuning.
When the two qubits are at resonance ($\Delta=0$), as during an iSWAP gate, the exchange term sets the quantizing field with corresponding dressed eigenstates
\begin{equation}
\ket{\tilde{1}}=\frac{\ket{10}+\ket{01}}{\sqrt{2}},
\qquad
\ket{\tilde{0}}=\frac{\ket{10}-\ket{01}}{\sqrt{2}},
\end{equation}
split by an energy $2g$. Thus, the subspace pseudo-spin in the coupled two-qubit system mirrors the spin-locking picture with the identification $\Omega\!\leftrightarrow\!2g$.

\paragraph{Dressed-state relaxation.}
Fluctuations of the differential detuning $\Delta$ (i.e.\ dephasing noise on $\omega_1$ and/or $\omega_2$) appear as a \emph{transverse} perturbation in the dressed basis and hence drive transitions between $\ket{\tilde{0}}$ and $\ket{\tilde{1}}$ when their spectral weight is at the dressed splitting $2g$. Writing the detuning noise as $\delta\Delta(t)$ with PSD
\begin{equation}
S_{\Delta}(\omega)=\int_{-\infty}^{\infty}\!dt\,e^{i\omega t}\,\big\langle\delta\Delta(t)\,\delta\Delta(0)\big\rangle,
\end{equation}
Bloch–Redfield theory gives the dressed-state relaxation rate $\Gamma_{1g} = \tfrac{1}{2}(\Gamma_1^{(1)}+\Gamma_1^{(2)})+\Gamma_{g}$, where $\Gamma_1^{(1,2)}$ is the $T_1$ relaxation rate for each qubit and  
\begin{equation}
\Gamma_g=\frac{1}{2}\,S_{\Delta}(2g) \;
\label{eq:Gg}
\end{equation}
is the extra decay term induced by $\delta\Delta(t)$ noise. The mechanism of $\Gamma_g$ is fully analogous to the $T_{1\rho}$ relaxation and is the focus of this study. Note that for the same noise source, $\Gamma_g=\Gamma_{\Omega}$ if $2g=\Omega$.

\paragraph{Contribution to gate error.}
In the classical limit ($k_B T\gg\hbar \, 2g$), the $\Gamma_g$ process can be modeled as a bit-flip channel defined in the dressed basis given by two equal-rate Lindbladians (see Supplemental Material \cite{SM} for more information),
\begin{equation}\label{eq:disspator}
\hat{L}_1 = \sqrt{\frac{\Gamma_{g}}{2}} \ket{\tilde{1}}\!\bra{\tilde{0}} \;,
\qquad
\hat{L}_2 = \sqrt{\frac{\Gamma_{g}}{2}} \ket{\tilde{0}}\!\bra{\tilde{1}}\;.
\end{equation}
The associated reduction in average gate fidelity under a weak Markovian error generator with jump operator $\hat{L}_j$ can be written as~\cite{universal_2022}
\begin{eqnarray}
\epsilon_{\mathrm{avg}}(\hat{L}_j)&=&\frac{d-1}{d}\,\mathrm{Tr}\!\left[\hat{L}_j^{\dagger}\hat{L}_j\right]\nonumber\\
&-&\frac{1}{d^2(d+1)}\sum_{i=1}^{d^2-1}\mathrm{Tr}\!\left[\hat{L}_j^{\dagger}\,\hat{f}_i\,\hat{L}_j\,\hat{f}_i\right],
\end{eqnarray}
where $d=2^N$ is the Hilbert-space dimension for $N$ qubits and $\{\hat{f}_i\}$ is any traceless operator basis. The total average error is then $\epsilon_{\text{avg}} = \sum_j\epsilon_{\text{avg}}(\hat{L}_j)$. Specializing to a two-qubit gate ($d=4$) of duration $\tau$ with dressed-state relaxation at rate $\Gamma_g$, the errors are
\begin{equation}
\epsilon_{\mathrm{avg}}=\frac{1}{5}\,\Gamma_g\,\tau \;,
\qquad
\epsilon=\frac{1}{4}\,\Gamma_g\,\tau \;,
\label{eq:errors-linear}
\end{equation}
to leading order in $\Gamma_g\tau\ll 1$, where $\epsilon = \frac{1-1/d^2}{1-1/d}\epsilon_{\mathrm{avg}}$ is the Pauli error.

\begin{figure}[t!]
\centering
\includegraphics[scale=1]{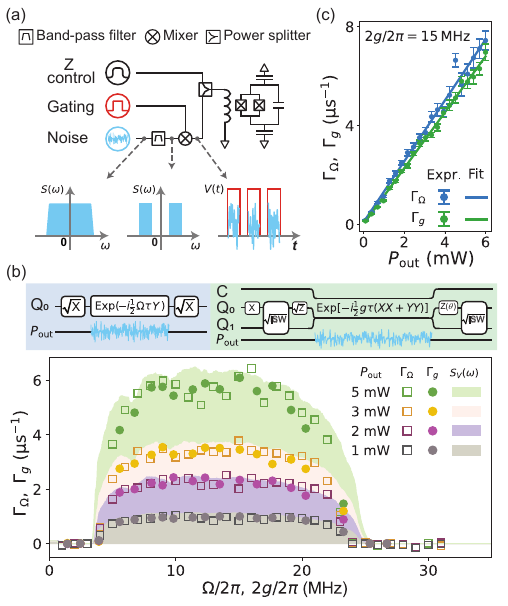}
\caption{\label{fig:gm1}
(a) Noise generation scheme. A Gaussian noise with $\sim$90~MHz bandwidth generated from an arbitrary waveform generator is first sent through a bandpass filter with a passband from $\rm 5\,MHz$ to $\rm 20\,MHz$, then envelope-mixed with a gating signal for selectively turning on the noise in specific time windows, and finally combined with regular flux control signal before being sent to the qubit.
(b) The measured $\Gamma_{\Omega}$ (squares) and $\Gamma_g$ (dots) extracted from the single-qubit spin-locking experiment (top-left panel) and the two-qubit dressed-state relaxation experiment (top-right panel), respectively, under different injected noise power $P_{\mathrm{out}}$.
The shaded areas are the noise spectrum $S_V(\omega)$ measured with a spectrum analyzer and converted to qubit detuning fluctuations. Note that the noise is turned on only during the locking or interaction period.
In the measurement protocol of the two-qubit dressed-state relaxation, we use a combination of single-qubit gates and a pre-calibrated square-root-of-iSWAP gate to prepare the qubits in the $\ket{\tilde{1}}=\frac{1}{\sqrt{2}}(\ket{10}+\ket{01})$ state. The locking period (XX+YY interaction) is followed by a phase gate with varying phase $\theta$ in order to extract the spin polarization in a robust way against any imperfect locking condition.
Note that the noise is only turned on during the locking period by the use of the gating signal to avoid disrupting state preparation and measurement.
(c) $\Gamma_{\Omega}$ (blue) and $\Gamma_g$ (green) versus $P_{\mathrm{out}}$ at a fixed Rabi frequency $\Omega/2\pi= 2g/2\pi$ = 15\,MHz from which we extract the rate-to-power ratios $\Gamma_{\Omega} / P_{\mathrm{out}} = 1.23\,(\mathrm{mW\cdot\mu s})^{-1}$ and $\Gamma_{g} / P_{\mathrm{out}} = 1.13\,(\mathrm{mW\cdot\mu s})^{-1}$. }
\end{figure}

\emph{Experimental demonstration.} 
To verify the above findings, we perform a set of experiments by injecting engineered noise onto superconducting transmon qubits and measuring the above dissipation processes. As shown in the schematic diagram in Fig.\,\ref{fig:gm1}(a), we leverage the built-in Gaussian noise generator in a commercial arbitrary waveform generator (AWG) together with a bandpass filter to create a tailored noise spectrum constrained within a 5-20\,MHz window. The noise signal is further mixed with a gating signal to ensure that the noise is switched on only in selected time windows. This process tailors the noise signal in both the frequency and time domains for later use. Finally, the processed noise signal is further combined with the regular flux control signal for biasing a transmon qubit. 

We verify the relation in Eq.\,(\ref{eq:Gg}) by comparing the measured $T_{1\rho}$ and $T_{1g}$ relaxation rates with the same added noise.
As demonstrated in Ref.~\cite{spinlocking2013}, the $T_{1\rho}$ noise spectroscopy technique utilizes the spin-locking sequence --- by preparing the qubit in the superposition state $\frac{1}{\sqrt{2}}(\ket{0}\pm\ket{1})$, locking it with a parallel drive field, and measuring the relaxation of pseudo-spin polarization --- probes the noise spectral densities at the Rabi frequencies which are tunable by microwave drive amplitude.
As the two-qubit analog of the single-qubit spin-locking experiment, the dressed-state relaxation experiment can probe the noise spectral densities at the same frequency provided that $2g=\Omega$.
Here, we use precalibrated single-qubit gates and square-root-of-iSWAP gate to prepare the two qubits into the $\ket{\tilde{1}}=\frac{1}{\sqrt{2}}(\ket{10}+\ket{01})$ state. During the locking period, the two transmon qubits are frequency-tuned into resonance while their exchange coupling, the $\ket{01}\!\bra{10}+\ket{10}\!\bra{01}$ term, is turned on by adjusting the tunable coupler~\cite{yan_tunable_2018}, so that the coupling field locks the subspace pseudo-spin. The coupling strength $g$ depends on the coupler frequency during interaction, and is extracted from the excitation swapping experiment \cite{SM}.  
The experimental protocols for both cases are shown in Fig.\,\ref{fig:gm1}(b).

The measured $\Gamma_{\Omega}$ and $\Gamma_g$ rates with different noise power $P_{\rm out}$ are shown in Fig.\,\ref{fig:gm1}(b). Plotted are the rates caused by added noise with the baseline rate subtracted, i.e.\ $\Gamma_{\Omega}(P_{\rm out})-\Gamma_{\Omega}(P_{\rm out}=0)$. Remarkably, $\Gamma_{\Omega}$ and $\Gamma_g$ agree with each other, and they both display a clear 5-20\,MHz pass-band with consistent magnitudes that scale with the injected noise power. This confirms that the added $T_{1g}$ relaxation is attributed to noise at frequency $2g$, fully analogous to that in the case of the $T_{1\rho}$ relaxation.
We note that near the edges of the passband, the measured relaxation rates are lower than expected from the noise spectra.
This is because the finite-duration pulse acts as an effective filter with a bandwidth on the order of $1/\tau$, centered around $\Omega$ or $2g$.
As a result, the noise power is only partially sampled at the passband edges, leading to reduced relaxation rates.

We also confirm the linear dependence on noise power by measuring both rates at a same frequency ($2g/2\pi=\Omega/2\pi=15$\,MHz) as shown in Fig.\,\ref{fig:gm1}(c), which yields approximately same slopes, $\partial\Gamma_{\Omega}/\partial P_{\rm out} = 1.23~{\rm (mW\cdot \mu s)^{-1}}$ and
$\partial\Gamma_{g}/\partial P_{\rm out} = 1.13\,{\rm (mW\cdot \mu s)^{-1}}$, validating the prefactors in Eq.\,(\ref{eq:Gg}). 
The value of $\Gamma_{g}$ being slightly smaller than $\Gamma_{\Omega}$ may be due to imperfect state locking.

\begin{figure}[thbp]
\centering
\includegraphics[scale=1]{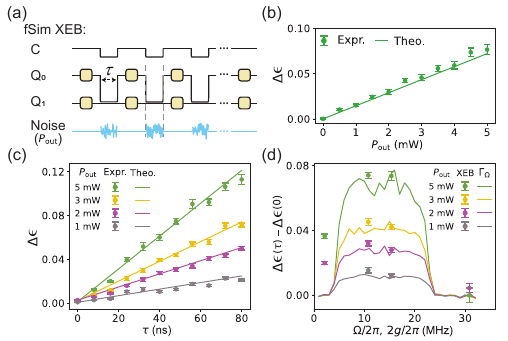}
\caption{\label{fig:gmg}(a) Circuit schematic for verifying the fidelity reduction characterized by $\Gamma_g$ under the application of an fSim gate. In the XEB circuit, the noise injection is timed to coincide with the two-qubit gate and only applied on one qubit. 
(b) $\Delta \epsilon$ as a function of input noise power. Green dots represent $\Delta\epsilon$ from XEB experiments at a coupling strength of 7.69\,MHz, with the solid green line showing the theoretical prediction.
(c) $\Delta \epsilon$ over interaction time $\tau$ from XEB experiments at input noise powers of 5\,mW (green dotted), 3\,mW (yellow dotted), 2\,mW (magenta dotted), and 1\,mW (gray dotted). The solid lines denote theoretical results.
(d) $\Delta\epsilon(P_{\rm out},\tau)-\Delta\epsilon(P_{\rm out},0)$ with the interaction duration of $\rm \tau=48\,ns$ as functions of Rabi frequency $\Omega$ and coupling strength $g$, respectively, under different input noise powers. The solid lines indicate the gate error $\Delta\epsilon$ induced by the measured $\Gamma_{\Omega}$, and the dots indicate the gate error measured by the XEB experiments.
}
\end{figure}

We next examine how such noise affects the performance of two-qubit entangling operations.
We use the same XX+YY exchange interaction described above as the generator of entangling gates, which naturally yields the $\mathrm{fSim}$ gate
\begin{equation}
\mathrm{fSim}(\theta, \phi) =
\begin{bmatrix}
1 & 0 & 0 & 0 \\
0 & \cos\theta & -i\sin\theta & 0 \\
0 & -i\sin\theta & \cos\theta & 0 \\
0 & 0 & 0 & e^{-i\phi}
\end{bmatrix},
\end{equation}
where $\theta$ is the swapping angle and $\phi$ is the controlled phase, both of which are learned from cross-entropy benchmarking (XEB).
For a given $\mathrm{fSim}$ gate, we perform the XEB protocol with and without noise injection (Fig.\,\ref{fig:gmg}(a)) and calculate their difference as the change in the Pauli error rate,
$\Delta\epsilon = \Delta\epsilon(P_{\rm out}, \tau) = \epsilon(P_{\rm out}, \tau) - \epsilon(0, \tau)$,
for a given noise power $P_{\rm out}$ and interaction time $\tau$.

We first study how the measured gate error depends on the noise power.
In this experiment, we fix the coupling strength to $g/2\pi = 7.69\,\mathrm{MHz}$, such that $2g$ lies within the filter passband, and use a gate duration consisting of a 48-ns interaction time with an additional 12-ns rise and fall, resulting in a unitary close to an iSWAP gate.
As shown in Fig.\,\ref{fig:gmg}(b), the measured increase in gate error (green dots) agrees well with the theoretical prediction from Eq.\,(\ref{eq:errors-linear}) (green line), using the independently calibrated $\Gamma_g$--$P_{\rm out}$ relation obtained in Fig.\,\ref{fig:gm1}(c).
This agreement demonstrates that the dressed-state relaxation induced by the injected noise quantitatively accounts for the observed increase in gate error.

We further investigate the dependence of the gate error on the interaction time $\tau$, which is the second key parameter in Eq.\,(\ref{eq:errors-linear}).
As shown in Fig.\,\ref{fig:gmg}(c), the measured error rates scale linearly with $\tau$ for different noise power settings, in good agreement with the theoretical expectation.

Finally, we examine how the gate error depends on the coupling strength $g$.
As shown in Fig.\,\ref{fig:gmg}(d), when $2g$ lies within the filter passband, the measured error rates agree well with the theoretical prediction across different noise powers.
The noise-induced error consistently vanishes around 30\,MHz, which is well above the passband.
By contrast, at the low-frequency end near 2\,MHz, the gate error remains appreciable.
This behavior arises in part because 2\,MHz lies closer to the passband edge, such that the noise is partially sampled during the finite gate duration, and in part because the dephasing noise, when strong enough, can still perturb the dressed-state pseudo-spin, albeit the noise is orthogonal to it (see supplement for details).
As a result, a longitudinal field at 2\,MHz produces a much stronger effect than one at 30\,MHz, even though the noise spectral density at both frequencies is negligible.
These observations indicate that stronger coupling strengths enhance the robustness of such two-qubit gates against dephasing.

\emph{Discussion} ---
Our results reveal that non-white noise with frequency components near the dressed-state splitting can decisively limit the performance of entangling operations.
The relevant figure of merit is the local power spectral density of detuning or dephasing noise, $S_{\Delta}(\omega)$, evaluated at the exchange-defined splitting.
This quantity directly determines the dressed-state relaxation rate $\Gamma_g$, although for finite-duration gates the effective error reflects a weighted average of the noise spectrum over a bandwidth set by the inverse gate time.

From a hardware perspective, suppressing electronics- and device-originated noise in the vicinity of $2g$ is therefore critical.
In typical superconducting quantum processors, the coupling strength is engineered to be $g/2\pi \sim 10\,\mathrm{MHz}$, corresponding to gate durations of tens of nanoseconds.
Unfiltered reference-clock feedthrough, mixer leakage, or control-line spurs in this frequency band can measurably increase $\Gamma_g$ and degrade gate fidelity.
It is thus essential to monitor and mitigate noise in this spectral window.
In addition, environmental features such as spurious modes and microscopic two-level defects can further contribute to noise near the dressed-state splitting.

Beyond mitigation, the dressed-state relaxation mechanism provides a spectroscopic tool.
When the coupling $g$ is tunable, it enables a narrowband probe of $S_{\Delta}(\omega)$ in frequency ranges that are difficult to access with conventional spin-locking techniques—for example, when the achievable Rabi frequency is limited by available drive power, or when continuous driving perturbs the system or its environment.
Combined use of dressed-state relaxation and spin-locking spectroscopy therefore offers a powerful means of cross-validating noise spectra.

Dressed-state relaxation also plays a central role in dual-rail encodings, where the logical states coincide with the dressed eigenstates.
Because the frequency fluctuations of the individual qubits are largely uncorrelated, they combine to form effective detuning noise, setting a fundamental limit on the logical energy-relaxation time~\cite{shim_2015-semiconductor-inspired,campbell_2020-universal,kubica_erasure_2023,teoh_dual-rail_2023,levine_2024-demonstrating,huang_logical_2025,chou_2024-superconducting}.
Similar considerations apply to singlet–triplet encodings in semiconductor quantum dots~\cite{johnson_tripletsinglet_2005,PhysRevLett.110.146804,connors_charge-noise_2022}, where an exchange-defined splitting likewise renders the logical qubit sensitive to detuning noise at the activated frequency.

\begin{acknowledgments}
The authors would like to thank Jiheng Duan and Xiaofei Liu for fruitful discussions. This work was supported by National Natural Science Foundation of China (Grants No.\,12404558, No.\,12322413, No.\,92476206, and No.\,92365206), Beijing Natural Science Foundation (Grants No.\,JQ25014), Innovation Program for Quantum Science and Technology (Grants No.\,2021ZD0301802).
\end{acknowledgments}

\bibliography{reference}

\newpage   
\clearpage 
\onecolumngrid

\begin{center}
\title{Supplemental Material for ``Dressed-state relaxation in coupled qubits as a source of two-qubit gate errors"}
\end{center}

\section{Derivation of Lindblad master equation for $T_{1\rho}$ and $T_{1g}$ processes}

This section derives the Lindblad master equations for the $T_{1\rho}$ and $T_{1g}$ decoherence processes in driven and correlated qubit systems. Starting from the general Lindblad form, we specialize it to obtain the master equation for $T_{1\rho}$, describing relaxation in a continuously driven rotating frame, and for $T_{1g}$, governing coherence decay between entangled states. Each derivation details the corresponding dissipators, their physical meaning, and the resulting density matrix dynamics.

\subsection{General derivation of Lindblad master equation \cite{breuer2002theory}}

Consider a quantum system $S$ weakly coupled to a reservoir $B$. The total Hamiltonian is
\begin{equation}
  H = H_S + H_B + H_I,
\end{equation}
where $H_S$ and $H_B$ are the free Hamiltonians of the system and reservoir, and $H_I$ is the interaction.

Define the interaction picture with respect to $H_0 = H_S + H_B$, there is $H_I(t) = e^{iH_0 t} H_I e^{-iH_0 t}$.

In the Schr{\"o}dinger picture, the density matrix $\rho_{\text{Sch}}(t)$ satisfies the von Neumann equation:
\begin{equation}
\frac{d}{dt} \rho_{\text{Sch}}(t) = -i[H, \rho_{\text{Sch}}(t)] \quad (\hbar=1).
\end{equation}

The interaction picture density matrix is defined as:
$\rho(t) = e^{iH_0 t} \, \rho_{\text{Sch}}(t) \, e^{-iH_0 t}$. The equation of motion for $\rho(t)$ becomes
$\frac{d}{dt}\rho(t) = -i [H_I(t), \rho(t)]$, integrating formally, we obtain
\begin{equation}
\rho(t) = \rho(0) - i\int_0^t dt' [H_I(t'), \rho(t')].
\label{EOM}
\end{equation}

Differentiating again and substituting into Eq.\,(\ref{EOM}), there is
\begin{align}
\frac{d}{dt}\rho(t) &= -i[H_I(t), \rho(0)] - \int_0^t dt' [H_I(t), [H_I(t'), \rho(t')]].
\end{align}

Define $\rho_S(t) = \operatorname{Tr}_B \rho(t)$. Take partial trace over $B$, we have
\begin{equation}
\frac{d}{dt}\rho_S(t) = -i \operatorname{Tr}_B[H_I(t), \rho(0)] - \int_0^t dt' \operatorname{Tr}_B[H_I(t), [H_I(t'), \rho(t')]].
\end{equation}
Assume the initial state is factorized and the reservoir is in thermal equilibrium
$\rho(0) = \rho_S(0) \otimes \rho_B$ and $\quad \rho_B = \frac{e^{-\beta H_B}}{Z}$,
where $Z = \operatorname{Tr}_B\left[ e^{-\beta H_B} \right]$ is a normalization constant ensuring that $\operatorname{Tr}_B(\rho_B) = 1$.

Typically, $\operatorname{Tr}_B[H_I(t) \rho_B] = 0$ holds because the interaction Hamiltonian $H_I$ is usually a tensor product of system and reservoir operators, $H_I = S \otimes R$. Since the reservoir is in thermal equilibrium $\rho_B$, the thermal average of typical reservoir operators vanishes, $\operatorname{Tr}_B[R \rho_B] = 0$, especially for linear couplings like $R = \sum_j (g_j b_j + g_j^* b_j^\dagger)$ where $b_j$ and $b_j^\dagger$ are annihilation and creation operators. This ensures no first-order coherent driving from the bath and simplifies the master equation derivation. So there is
\begin{equation}
\frac{d}{dt}\rho_S(t) = - \int_0^t dt' \operatorname{Tr}_B[H_I(t), [H_I(t'), \rho(t')]].
\end{equation}

Assume the coupling is weak and the reservoir is large (Born Approximation), so $\rho(t) \approx \rho_S(t) \otimes \rho_B$ for all $t$, thus there is
\begin{equation}
\frac{d}{dt}\rho_S(t) = - \int_0^t dt' \operatorname{Tr}_B[H_I(t), [H_I(t'), \rho_S(t') \otimes \rho_B]].
\end{equation}

Assume the system dynamics only depend on the current state (Markov Approximation) $\rho_S(t)$ (replace $\rho_S(t') \to \rho_S(t)$), there is
\begin{equation}
\frac{d}{dt}\rho_S(t) = - \int_0^t dt' \operatorname{Tr}_B[H_I(t), [H_I(t'), \rho_S(t) \otimes \rho_B]].
\end{equation}
This is the Redfield equation.

Let $\tau = t - t'$, then,
\begin{equation}
\frac{d}{dt}\rho_S(t) = - \int_0^t d\tau \, \operatorname{Tr}_B[H_I(t), [H_I(t-\tau), \rho_S(t) \otimes \rho_B]].
\end{equation}
Assume the reservoir correlation time $\tau_B$ is short compared to the system evolution time, so we can extend the upper limit to infinity
\begin{equation}
\frac{d}{dt}\rho_S(t) = - \int_0^\infty d\tau \, \operatorname{Tr}_B[H_I(t), [H_I(t-\tau), \rho_S(t) \otimes \rho_B]].
\end{equation}

Assume the interaction Hamiltonian can be expressed as a sum of tensor products of Hermitian operators
$H_I = \sum_\alpha S_\alpha \otimes R_\alpha$,
where $S_\alpha$ are system operators and $R_\alpha$ are reservoir operators, both Hermitian. Substituting into the Redfield equation and expanding gives
\begin{align}
\frac{d\rho_S}{dt} &= \sum_{\alpha,\beta} \int_0^\infty d\tau \, C_{\alpha\beta}(\tau) \left( S_\beta(t-\tau) \rho_S S_\alpha(t) - S_\alpha(t) S_\beta(t-\tau) \rho_S \right) + \text{H.c.},
\end{align}
where H.c. stands for Hermitian conjugate of the preceding terms and $C_{\alpha\beta}(\tau) = \langle R_\alpha(\tau) R_\beta(0) \rangle$ is the reservoir correlation function.

Expand system operators in the eigenbasis of $H_S$, $S_\alpha(t) = \sum_\omega e^{i\omega t} S_\alpha(\omega)$,
where $S_\alpha(\omega)$ are transition operators satisfying $[H_S, S_\alpha(\omega)] = \omega S_\alpha(\omega)$.
For Hermitian $S_\alpha(t)$, we have $S_\alpha^\dagger(\omega) = S_\alpha(-\omega)$.
Keep only resonant terms where frequencies sum to zero ($\omega + \omega' = 0$). This eliminates rapidly oscillating terms and gives
\begin{equation}
S_\beta(t-\tau) \rho_S S_\alpha(t) \to e^{-i\omega\tau} S_\beta(\omega) \rho_S S_\alpha(-\omega) = e^{-i\omega\tau} S_\beta(\omega) \rho_S S^\dagger_\alpha(\omega),
\end{equation}
and similarly for other terms.

Define the spectral density as
\begin{equation}
\Gamma_{\alpha\beta}(\omega) = \int_0^\infty d\tau \, e^{-i\omega\tau} C_{\alpha\beta}(\tau),
\end{equation}
and decompose into Hermitian and anti-Hermitian parts, there is
\begin{equation}
\Gamma_{\alpha\beta}(\omega) = \frac{1}{2}\gamma_{\alpha\beta}(\omega) + i \Delta_{\alpha\beta}(\omega),
\end{equation}
where $\gamma_{\alpha\beta}(\omega) = \Gamma_{\alpha\beta}(\omega) + \Gamma_{\beta\alpha}^*(\omega)$ and $\quad
\Delta_{\alpha\beta}(\omega) = \frac{1}{2i}[\Gamma_{\alpha\beta}(\omega) - \Gamma_{\beta\alpha}^*(\omega)]$.

After RWA and using reservoir symmetry for thermal equilibrium, we obtain
\begin{equation}
\frac{d\rho_S}{dt} = \sum_{\omega} \sum_{\alpha,\beta} \left[ \Gamma_{\alpha\beta}(\omega) \left( S_\beta(\omega) \rho_S S_\alpha^\dagger(\omega) - S_\alpha^\dagger(\omega) S_\beta(\omega) \rho_S \right) + \text{H.c.} \right].
\end{equation}

The dissipative part (from $\gamma_{\alpha\beta}$) becomes
\begin{equation}
\frac{d\rho_S}{dt}\bigg|_{\text{diss}} = \sum_{\omega} \sum_{\alpha,\beta} \frac{1}{2}\gamma_{\alpha\beta}(\omega) \left[ S_\beta(\omega) \rho_S S_\alpha^\dagger(\omega) + S_\alpha(\omega) \rho_S S_\beta^\dagger(\omega) - S_\alpha^\dagger(\omega) S_\beta(\omega) \rho_S - \rho_S S_\beta^\dagger(\omega) S_\alpha(\omega) \right].
\label{diss}
\end{equation}

The coherent part (from $\Delta_{\alpha\beta}$) gives the Lamb shift Hamiltonian as $H_{\text{LS}} = \sum_{\omega} \sum_{\alpha,\beta} \Delta_{\alpha\beta}(\omega) S_\alpha^\dagger(\omega) S_\beta(\omega)$.
Since $\gamma_{\alpha\beta}(\omega)$ is Hermitian positive semidefinite, it can be diagonalized as $\gamma_{\alpha\beta}(\omega) = \sum_k U_{\alpha k}(\omega) \gamma_k(\omega) U_{\beta k}^*(\omega)$. 
Define new Lindblad operators as $L_k(\omega) = \sum_\alpha U_{\alpha k}^*(\omega) S_\alpha(\omega)$, and substituting into Eq.\,(\ref{diss}) the dissipative part becomes standard Lindblad form, which is
\begin{equation}
\frac{d\rho_S}{dt}\bigg|_{\text{diss}} = \sum_{\omega,k} \gamma_k(\omega) \left( L_k(\omega) \rho_S L_k^\dagger(\omega) - \frac{1}{2} \{ L_k^\dagger(\omega) L_k(\omega), \rho_S \} \right).
\end{equation}
Combining with the coherent part, we obtain the complete Lindblad master equation:
\begin{equation}
\frac{d}{dt}\rho_S(t) = -i [H_S + H_{\text{LS}}, \rho_S(t)] + \sum_k \gamma_k \left( L_k \rho_S(t) L_k^\dagger - \frac{1}{2} \{ L_k^\dagger L_k, \rho_S(t) \} \right),
\end{equation}
where the sum over $k$ now includes all frequency modes.

After expanding in system eigenbasis and performing the secular approximation (keeping only resonant terms), one obtains the Lindblad form:
\begin{equation}
\frac{d}{dt}\rho_S(t) = -i [H_{\text{eff}}, \rho_S(t)] + \sum_k \gamma_k \left( L_k \rho_S(t) L_k^\dagger - \frac{1}{2}\{L_k^\dagger L_k, \rho_S(t)\} \right),
\end{equation}
where $H_{\text{eff}}$ is a Lamb-shifted Hamiltonian, $L_k$ are jump operators, and $\gamma_k \ge 0$ are decay rates.

\subsection{Lindblad master equation for $T_{1\rho}$ process}

For $T_{1\rho}$ process, only consider the noise transverse to the
driving field at frequency $\Omega$, the original Hamiltonian is 
\begin{equation}
H=\frac{\omega}{2}\sigma_{z}+\Omega\cos(\omega t)\sigma_{x}+\sum_{j}\omega_{j}b_{j}^{\dagger}b_{j}+\sigma_{z}\sum_{j}(g_{j}b_{j}^{\dagger}+g_{j}^{\ast}b_{j}).
\end{equation}

Applying the rotating wave approximation with the transform unitary $U(t)=e^{-i\frac{\omega}{2}\sigma_{z}t-i\sum_{j}\omega_{j}b_{j}^{\dagger}b_{j}t}$, there is
\begin{equation}
H^{\prime}	=	\frac{\Omega}{2}\sigma_{x}+\sigma_{z}\sum_{j}(g_{j}b_{j}^{\dagger}e^{i\omega_{j}t}+g_{j}^{\ast}b_{j}e^{-i\omega_{j}t}).
\end{equation}

In the basis of $|\tilde{1}\rangle=\frac{1}{\sqrt{2}}(|0\rangle+|1\rangle)$ and $|\tilde{0}\rangle=\frac{1}{\sqrt{2}}(|0\rangle-|1\rangle)$, there are $\sigma_{z}=|\tilde{0}\rangle\langle\tilde{1}|+|\tilde{1}\rangle\langle\tilde{0}|$
and
\begin{align}
&e^{i\frac{\Omega}{2}\sigma_{x}t}|\tilde{0}\rangle\langle\tilde{1}|e^{-i\frac{\Omega}{2}\sigma_{x}t}=e^{-i\Omega t}|\tilde{0}\rangle\langle\tilde{1}|\nonumber\\
&e^{i\frac{\Omega}{2}\sigma_{x}t}|\tilde{1}\rangle\langle\tilde{0}|e^{-i\frac{\Omega}{2}\sigma_{x}t}=e^{i\Omega t}|\tilde{1}\rangle\langle\tilde{0}|.
\end{align}

Applying the rotating wave approximation with the transform unitary $U(t)=e^{-i\frac{\Omega}{2}\sigma_{x}t}$, and only retain the terms with $\Omega=\omega_{j}$, the interaction Hamiltonian becomes
\begin{align}
H_{I}&=	(|\tilde{0}\rangle\langle\tilde{1}|e^{-i\Omega t}+|\tilde{1}\rangle\langle\tilde{0}|e^{i\Omega t})\sum_{j}(g_{j}b_{j}^{\dagger}e^{i\omega_{j}t}+g_{j}^{\ast}b_{j}e^{-i\omega_{j}t})\nonumber\\
&\approx	g_{j}|\tilde{0}\rangle\langle\tilde{1}|b_{j,\omega_{j}=\Omega}^{\dagger}+g_{j}^{\ast}|\tilde{1}\rangle\langle\tilde{0}|b_{j,\omega_{j}=\Omega}.
\end{align}

Define $S(t)=|\tilde{0}\rangle\langle\tilde{1}|e^{-i\Omega t}+|\tilde{1}\rangle\langle\tilde{0}|e^{i\Omega t}=\tilde{\sigma}_{-}e^{-i\Omega t}+\tilde{\sigma}_{+}e^{i\Omega t}$ and $R(t)=\sum_{j}(g_{j}b_{j}^{\dagger}e^{i\omega_{j}t}+g_{j}^{\ast}b_{j}e^{-i\omega_{j}t})$, there are $H_{I}=S(t)\otimes R(t)$ and
$S(t)=e^{-i\Omega t}S(\Omega)+e^{i\Omega t}S(-\Omega)$,
\begin{align}
C(\tau)	&=	\langle R(\tau)R(0)\rangle\nonumber\\
&=\langle\sum_{j}(g_{j}b_{j}^{\dagger}e^{i\omega_{j}\tau}+g_{j}^{\ast}b_{j}e^{-i\omega_{j}\tau})\sum_{i}(g_{j}b_{i}^{\dagger}+g_{j}^{\ast}b_{i})\rangle\nonumber\\
&=\langle\sum_{j}(|g_{j}|^{2}b_{j}^{\dagger}b_{j}e^{i\omega_{j}\tau}+|g_{j}|^{2}b_{j}b_{j}^{\dagger}e^{-i\omega_{j}\tau})\rangle\nonumber\\
&=\sum_{j}\{|g_{j}|^{2}n_{th}(\omega_{j})e^{i\omega_{j}\tau}+|g_{j}|^{2}[n_{th}(\omega_{j})+1]e^{-i\omega_{j}\tau}\},
\end{align}
where $n_{th}(\omega_{j})=\frac{1}{e^{\beta\omega_{j}}-1}$ is the average paticle number in a state of thermal equilibtium, 
$\langle b_{i}^{\dagger}b_{j}^{\dagger}\rangle=0, \langle b_{i}b_{j}\rangle=0, \langle b_{i}^{\dagger}b_{j}\rangle_{i\neq j}=0$ and $\langle b_{i}b_{j}^{\dagger}\rangle_{i\neq j}=0$. Define the spectral density $J(\omega)=\sum_{j}|g_{j}|^{2}\delta(\omega-\omega_{j})$, and there is $\sum_{j}|g_{j}|^{2}f(\omega_{j})=\int_{0}^{\infty}d\omega J(\omega)f(\omega)$, applying this to $C(\tau)$, we obtain
\begin{equation}
C(\tau)=\int_{0}^{\infty}d\omega J(\omega)\{n_{th}(\omega)e^{i\omega\tau}+[n_{th}(\omega)+1]e^{-i\omega\tau}\}.
\end{equation}

The one-sided Fourior transform is
\begin{align}
\Gamma(\Omega)&=	\int_{0}^{\infty}d\tau e^{-i\Omega\tau}C(\tau)\nonumber\\
&=\int_{0}^{\infty}d\tau e^{-i\Omega\tau}\int_{0}^{\infty}d\omega J(\omega)\{n_{th}(\omega)e^{i\omega\tau}+[n_{th}(\omega)+1]e^{-i\omega\tau}\}\nonumber\\
&=\int_{0}^{\infty}J(\omega)n_{th}(\omega)\int_{0}^{\infty}e^{i(\omega-\Omega)\tau}d\tau d\omega\nonumber\\
&=\int_{0}^{\infty}J(\omega)n_{th}(\omega)[\pi\delta(\omega-\Omega)+i\mathcal{P}\frac{1}{\omega-\Omega}]d\omega\nonumber\\
&=\pi J(\Omega)n_{th}(\Omega)+i\mathcal{P}\int_{0}^{\infty}\frac{J(\omega)n_{th}(\Omega)}{\omega-\Omega}d\omega
\end{align}

and
\begin{align}
\Gamma(-\Omega)&=\pi J(\Omega)[n_{th}(\Omega)+1]+i\mathcal{P}\int_{0}^{\infty}\frac{J(\omega)[n_{th}(\omega)+1]}{\Omega-\omega}d\omega.
\end{align}

The dissipation rates are
\begin{align}
&\gamma_{+}=2\pi J(\Omega)n_{th}(\Omega)\nonumber\\
&\gamma_{-}=2\pi J(\Omega)[n_{th}(\Omega)+1].
\end{align}

Ignore the small amount of Lamb shift, the Lindblad equation is
\begin{align}
\frac{d\rho}{dt}=\gamma_{+}(\tilde{\sigma}_{+}\rho\tilde{\sigma}_{-}-\frac{1}{2}\{\tilde{\sigma}_{-}\tilde{\sigma}_{+},\rho\})+\gamma_{-}(\tilde{\sigma}_{-}\rho\tilde{\sigma}_{+}-\frac{1}{2}\{\tilde{\sigma}_{+}\tilde{\sigma}_{-},\rho\}),
\end{align}
and the Lindblad operators are $L_1=\tilde{\sigma}_{+}=|\tilde{1}\rangle\langle\tilde{0}|$, and $L_2=\tilde{\sigma}_{-}=|\tilde{0}\rangle\langle\tilde{1}|$.

\subsection{Lindblad master equation for $T_{1g}$ process}

The $T_{1g}$ process characterizes the relaxation of a system comprising two coupled two-level systems (qubits) subjected to a common dephasing bath when the qubits are resonantly tuned. This analysis begins with the full Hamiltonian describing the two qubits coupled to each other and to a bosonic bath via a collective dephasing term as
\begin{align}
H_{1g}=\frac{\omega_{0}}{2}\sigma_{0}^{z}+\frac{\omega_{1}}{2}\sigma_{1}^{z}+g(a_{0}^{\dagger}a_{1}+a_{1}^{\dagger}a_{0})+\sum_{j}\omega_{j}b_{j}^{\dagger}b_{j}+\sigma_{0}^{z}\sum(g_{j}b_{j}^{\dagger}+g_{j}^{\ast}b_{j}).
\end{align}

We consider the resonant condition $\omega_{0}=\omega_{1}$. To simplify the dynamics, we transform into a rotating frame defined by the unitary $U=e^{-i(\frac{\omega_{0}}{2}\sigma_{0}^{z}+\frac{\omega_{1}}{2}\sigma_{1}^{z})t-i\sum_{j}\omega_{j}b_{j}^{\dagger}b_{j}t}$. Applying the rotating wave approximation (RWA) eliminates the fast-oscillating terms, yielding the effective Hamiltonian in the interaction picture, which can be written as 
\begin{align}
H_{1g}^{\prime}=g(a_{0}^{\dagger}a_{1}+a_{1}^{\dagger}a_{0})+\sigma_{0}^{z}\sum_{j}(g_{j}b_{j}^{\dagger}e^{i\omega_{j}t}+g_{j}^{\ast}b_{j}e^{-i\omega_{j}t}).
\end{align}

It is advantageous to express the system in the basis of $|00\rangle$, $|\tilde{1}\rangle=\frac{1}{\sqrt{2}}(|01\rangle+|10\rangle)$, $|\tilde{0}\rangle=\frac{1}{\sqrt{2}}(|01\rangle-|10\rangle)$, and $|11\rangle$. In this basis, the operators take the following forms as
\begin{align}
&a_{0}^{\dagger}a_{1}+a_{1}^{\dagger}a_{0}=|\tilde{1}\rangle\langle\tilde{1}|-|\tilde{0}\rangle\langle\tilde{0}|\nonumber\\ &\sigma_{0}^{z}=|00\rangle\langle00|+|\tilde{1}\rangle\langle\tilde{0}|+|\tilde{0}\rangle\langle\tilde{1}|+|11\rangle\langle11|.
\end{align}

Substituting these expressions into $H_{1g}^{\prime}$ gives
\begin{align}
H_{1g}^{\prime}=\frac{2g}{2}(|\tilde{1}\rangle\langle\tilde{1}|-|\tilde{0}\rangle\langle\tilde{0}|)+(|00\rangle\langle00|+|\tilde{1}\rangle\langle\tilde{0}|+|\tilde{0}\rangle\langle\tilde{1}|-|11\rangle\langle11|)\sum_{j}(g_{j}b_{j}^{\dagger}e^{i\omega_{j}t}+g_{j}^{\ast}b_{j}e^{-i\omega_{j}t}).
\end{align}

We now move into a second interaction picture with respect to the swap coupling, using the transformation $U=e^{-ig(|\tilde{1}\rangle\langle\tilde{1}|-|\tilde{0}\rangle\langle\tilde{0}|)t}$, we obtain
\begin{align}
H_{I}	=	(|00\rangle\langle00|+|\tilde{1}\rangle\langle\tilde{0}|e^{i2gt}+|\tilde{0}\rangle\langle\tilde{1}|e^{-i2gt}-|11\rangle\langle11|)\sum_{j}(g_{j}b_{j}^{\dagger}e^{i\omega_{j}t}+g_{j}^{\ast}b_{j}e^{-i\omega_{j}t}).
\end{align}
Assuming the bath spectral density $J(\omega)=\sum_{j}|g_{j}|^{2}\delta(\omega-\omega_{j})$ is defined for $\omega>|\omega_{0}|$ and that the noise power is concentrated within a specific frequency band, we can apply a second RWA. 
This approximation—neglecting terms oscillating at frequencies far from resonance—potentially contributes to the discrepancy observed between experimental and theoretical results outside the filter passband, and warrants detailed investigation in future work.
Under this rotating-wave approximation, only those interaction terms that are nearly resonant with the system’s internal dynamics—specifically, the exchange between the states $|\tilde{1}\rangle$ and $|\tilde{0}\rangle$ at frequency $2g$—are retained, while rapidly oscillating off-resonant terms are averaged to zero. Consequently, the effective interaction Hamiltonian simplifies to the following form as
\begin{align}
H_{I}\approx(|\tilde{1}\rangle\langle\tilde{0}|e^{i2gt}+|\tilde{0}\rangle\langle\tilde{1}|e^{-i2gt})\sum_{j}(g_{j}b_{j}^{\dagger}e^{i\omega_{j}t}+g_{j}^{\ast}b_{j}e^{-i\omega_{j}t}).
\end{align}

The structure of this interaction Hamiltonian is formally equivalent to that of a driven two-level system coupled to a bath, analogous to the $T_{1\rho}$ process in a single qubit. Following the standard derivation of the Lindblad master equation under the Born-Markov approximation, we obtain the dissipative dynamics for the $T_{1g}$ process as
\begin{align}
\frac{d\rho}{dt}=\gamma_{+}(\tilde{\sigma}_{+}\rho\tilde{\sigma}_{-}-\frac{1}{2}\{\tilde{\sigma}_{-}\tilde{\sigma}_{+},\rho\})+\gamma_{-}(\tilde{\sigma}_{-}\rho\tilde{\sigma}_{+}-\frac{1}{2}\{\tilde{\sigma}_{+}\tilde{\sigma}_{-},\rho\}).
\end{align}
Here, $\tilde{\sigma}_{+}=|\tilde{1}\rangle\langle\tilde{0}|$ and $\tilde{\sigma}_{-}=|\tilde{0}\rangle\langle\tilde{1}|$ are the effective raising and lowering operators between the states $|\tilde{0}\rangle$ and $|\tilde{1}\rangle$. The transition rates are determined by the bath spectral density evaluated at the splitting frequency $2g$, which can be expressed as $\gamma_{+}=2\pi J(2g)n_{th}(2g)$ and $\gamma_{-}=2\pi J(2g)[n_{th}(2g)+1]$, where $n_{th}(\omega)$ is the thermal occupation number of the bath at frequency $2g$. This master equation describes the incoherent relaxation between the entangled states induced by the common dephasing noise.

\section{Derivation of the Relationship between $\Gamma_{1g}$ and $\Gamma_g$}

We consider the system's density matrix expressed in the basis formed by the Bell-like states $|\tilde{1}\rangle = \frac{1}{\sqrt{2}}(|01\rangle + |10\rangle)$, $|\tilde{0}\rangle = \frac{1}{\sqrt{2}}(|01\rangle - |10\rangle)$, and the ground state $|00\rangle$. In this representation, the relevant dissipative processes are described by the jump operators
\begin{align}
\hat{a}_{0}=|00\rangle\langle10|=\left[\begin{array}{ccc}
0 & 0 & 0\\
0 & 0 & 0\\
\frac{1}{\sqrt{2}} & -\frac{1}{\sqrt{2}} & 0
\end{array}\right],\;  \hat{a}_{1}=|00\rangle\langle01|=\left[\begin{array}{ccc}
0 & 0 & 0\\
0 & 0 & 0\\
\frac{1}{\sqrt{2}} & \frac{1}{\sqrt{2}} & 0
\end{array}\right],
\end{align}
which correspond to decay from the single-excitation states $|10\rangle$ and $|01\rangle$ into the ground state $|00\rangle$, with associated rates $\Gamma_{0}$ and $\Gamma_{1}$, respectively. Additionally, we include the dissipation process between the states $|\tilde{1}\rangle$ and $|\tilde{0}\rangle$. 
This is modeled by the operators $\hat{a}_{01}=|\tilde{0}\rangle\langle\tilde{1}|$ and $\hat{a}_{10}=|\tilde{1}\rangle\langle\tilde{0}|$.

In the ordered basis  ${|\tilde{1}\rangle, |\tilde{0}\rangle, |00\rangle}$ and assuming $T_1$ relaxation process  is dominated by the decay rate $\Gamma_{\downarrow}$ while neglecting the excitation rate $\Gamma_{\uparrow}$, the Lindblad master equation governing the system's dynamics is given by
\begin{align}
\left[\begin{array}{ccc}
\dot{\rho}_{00} & \dot{\rho}_{01} & \dot{\rho}_{02}\\
\dot{\rho}_{10} & \dot{\rho}_{11} & \dot{\rho}_{12}\\
\dot{\rho}_{20} & \dot{\rho}_{21} & \dot{\rho}_{22}
\end{array}\right]	&=	\Gamma_{1}^{(1)}(\hat{a}_{1}\rho\hat{a}_{1}^{\dagger}-\frac{1}{2}\{\hat{a}_{1}^{\dagger}\hat{a}_{1},\rho\})+\Gamma_{1}^{(0)}(\hat{a}_{0}\rho\hat{a}_{0}^{\dagger}-\frac{1}{2}\{\hat{a}_{0}^{\dagger}\hat{a}_{0},\rho\})\\
&+\frac{\Gamma_{g}}{2}(\hat{a}_{10}\rho\hat{a}_{10}^{\dagger}-\frac{1}{2}\{\hat{a}_{10}^{\dagger}\hat{a}_{10},\rho\})+\frac{\Gamma_{g}}{2}(\hat{a}_{01}\rho\hat{a}_{01}^{\dagger}-\frac{1}{2}\{\hat{a}_{01}^{\dagger}\hat{a}_{01},\rho\}).
\end{align}

By explicitly evaluating the terms of this master equation, we obtain the following differential equations for the diagonal matrix elements $\rho_{00}$ and $\rho_{11}$ as
\begin{align}
\dot{\rho}_{00}&=-\frac{\Gamma_{1}^{(0)}}{4}(2\rho_{00}-\rho_{01}-\rho_{10})-\frac{\Gamma_{1}^{(1)}}{4}(2\rho_{00}+\rho_{01}+\rho_{10})+\frac{\Gamma_{g}}{2}\rho_{11}-\frac{\Gamma_{g}}{2}\rho_{00}\\
\dot{\rho}_{11}&=-\frac{\Gamma_{1}^{(0)}}{4}(-\rho_{01}-\rho_{10}+2\rho_{11})-\frac{\Gamma_{1}^{(1)}}{4}(\rho_{01}+\rho_{10}+2\rho_{11})-\frac{\Gamma_{g}}{2}\rho_{11}+\frac{\Gamma_{g}}{2}\rho_{00}.
\end{align}

We are interested in the dynamics of the population difference between $|\tilde{1}\rangle$ and $|\tilde{0}\rangle$, which is given by $\rho_{00} - \rho_{11}$. Subtracting the second equation from the first yields
\begin{align}
\dot{\rho}_{00}-\dot{\rho}_{11}&=-\frac{\Gamma_{1}^{0}}{2}(\rho_{00}-\rho_{11})-\frac{\Gamma_{1}^{(1)}}{2}(\rho_{00}-\rho_{11})+\Gamma_{g}\rho_{11}-\Gamma_{g}\rho_{00}.
\end{align}

This expression can be simplified by grouping the terms proportional to $(\rho_{00} - \rho_{11})$ as
\begin{align}
\dot{\rho}_{00}-\dot{\rho}_{11}	=	-\frac{\Gamma_{1}^{(0)}+\Gamma_{1}^{(1)}}{2}(\rho_{00}-\rho_{11})-\Gamma_{g}(\rho_{00}-\rho_{11}).
\end{align}

We can now identify the total effective decay rate for the population difference. Factoring out $(\rho_{00}-\rho_{11})$ gives an equation of the form $\frac{d}{dt}(\rho_{00}-\rho_{11}) = -\Gamma_{1g} (\rho_{00}-\rho_{11})$, from which we directly read off the relationship between the total dissipative rate $\Gamma_{1g}$ and the constituent rates $\Gamma_{1g}=\frac{\Gamma_1^{(0)}+\Gamma_1^{(1)}}{2}+\Gamma_g$.
Thus, the total dissipation rate $\Gamma_{1g}$ governing the relaxation of the population difference between the Bell states is the sum of the average decay rate from the single-excitation subspace and the pure dephasing rate $\Gamma_g$ between the Bell states.

\section{Experimental setup}\label{append:exp_setup}

\subsection{Qubit information}

This experiment employs a superconducting quantum processor analogous to the architecture detailed in Ref.~\cite{chen2025efficient}. The system comprises two frequency-tunable transmon qubits, denoted as $\mathrm{Q}_0$ and $\mathrm{Q}_1$, which are capacitively coupled via an intermediate frequency-tunable coupler. The coupler's transition frequency is dynamically adjustable, enabling precise control over the effective coupling strength $g$ between the two qubits.

Key spectroscopic and coherence parameters for both qubits are summarized in Table~\ref{tab:table1}. Specifically, we list the maximum ($\omega^{\rm max}$) and idle ($\omega^{\rm idle}$) frequencies (corresponding to the $\ket{0} \rightarrow \ket{1}$ transition), the qubit anharmonicity $\alpha$, and the energy relaxation time $T_1$ and dephasing time $T_2$ measured at both frequency points. Here, $T_2$ is obtained via a Ramsey interferometry sequence and is therefore denoted as $T_{2R}$. The idle point represents the operating frequency where single-qubit gates are typically performed, whereas the maximum point corresponds to the upper limit of the qubit's frequency tuning range.

\begin{table}[h!]
\begin{threeparttable}
    \centering
    \begin{tabular}{|c|c|c|c|}
    \hline
        & $\rm Q_0$ & $\rm Q_1$ \\
        \colrule
        ~~$\omega^{\rm max}/2\pi$ (GHz)~~ & ~~4.29~~ & ~~4.31~~ \\

        ~~$\omega^{\rm idle}/2\pi$ (GHz)~~ & ~~4.19~~ & ~~4.28~~ \\
        
        ~~$\alpha/2\pi$ (MHz) ~~& ~~-192.0~~ & ~~-198.4~~\\
       
        ~~$T_1^{\rm max}$ ($\mu$s) ~~ & ~~51.7~~ & ~~67.77~~\\

        ~~$T_1^{\rm idle}$ ($\mu$s) ~~ & ~~66.6~~ & ~~99.8~~\\
        
        ~~$T_{2R}^{\rm max}$ ($\mu$s) ~~ & ~~26.6~~ & ~~9.17~~\\
        
        ~~$T_{2R}^{\rm idle}$ ($\mu$s) ~~ & ~~2.7~~ & ~~3.2~~\\
    \hline
    \end{tabular}
\end{threeparttable}
\caption{\label{tab:table1} Measured parameters for the two tunable transmon qubits used in the experiment.}
\end{table}

\subsection{Characterization of AWG output noise power}

\begin{figure}[tbhp]
\centering
\includegraphics[scale=1]{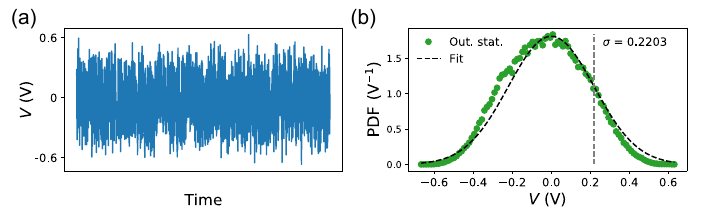}
\caption{\label{suppfig:Noise_power} Statistical analysis of the AWG output noise.
(a) The time-domain output signal of the AWG. (b) The probability density function (PDF) constructed from the voltage data shown in panel (a). The empirical distribution (green dots) is closely approximated by a Gaussian fit (black dashed line), confirming the stochastic nature of the output noise.
}
\end{figure}

The power of the additive output noise generated by the arbitrary waveform generator (AWG) is quantified through a direct time-domain measurement and subsequent statistical analysis.
To this end, the AWG voltage output, which is set to 1\,V, is connected to an oscilloscope with a matched impedance of $R = 50\,\Omega$. A continuous stream of the output signal is recorded, yielding a time-series dataset as displayed in Fig.\,\ref{suppfig:Noise_power}(a).

We then perform a statistical characterization of the acquired voltage samples. From the voltage time-series data, we directly estimate the empirical probability density function (PDF), which is plotted as discrete points (green dots) in Fig.\,\ref{suppfig:Noise_power}(b) with voltage $V$ on the horizontal axis and probability density on the vertical axis. The distribution exhibits a symmetric, bell-shaped profile, which we fit to a Gaussian function of the form $\frac{1}{\sqrt{2\pi \sigma^2}} \exp\left(-\frac{V^2}{2\sigma^2}\right)$. The excellent agreement between the data and the fit (black dashed line) validates that the output noise voltage follows a zero-mean Gaussian stochastic process. The fit yields a standard deviation of $\sigma = 0.2203\,{\rm V}$, which fully characterizes the noise amplitude.

From this statistical model, a direct measure of the noise power delivered into a matched load is derived. For a zero-mean Gaussian variable, the expectation value $\mathbb{E}[V^2]$ is given by the variance:
\begin{equation}
\mathbb{E}[{V^2}] = \int_{-\infty}^{\infty} V^2  \frac{1}{\sqrt{2\pi \sigma^2}} e^{-\frac{V^2}{2\sigma^2}} dV = \sigma^2.
\end{equation}
The corresponding average output noise power $P_{\rm out}$ is therefore:
\begin{equation}
P_{\rm out} = \frac{\mathbb{E}[V^2]}{R} = \frac{\sigma^2}{R}.
\end{equation}
Substituting the measured $\sigma = 0.2203\,{\rm V}$ and $R = 50\, \Omega$ yields $P_{\rm out} \approx 1\, {\rm mW}$ for the nominal 1\,V AWG setting.

This procedure provides a calibration curve linking the configured AWG voltage level to its effective output noise power. We systematically applied this characterization method to determine $P_{\mathrm{out}}$ for all relevant voltage settings used in the subsequent experiments.

\subsection{Experimental configuration for classical noise injection}

\begin{figure}[tbhp]
\centering
\includegraphics[scale=1]{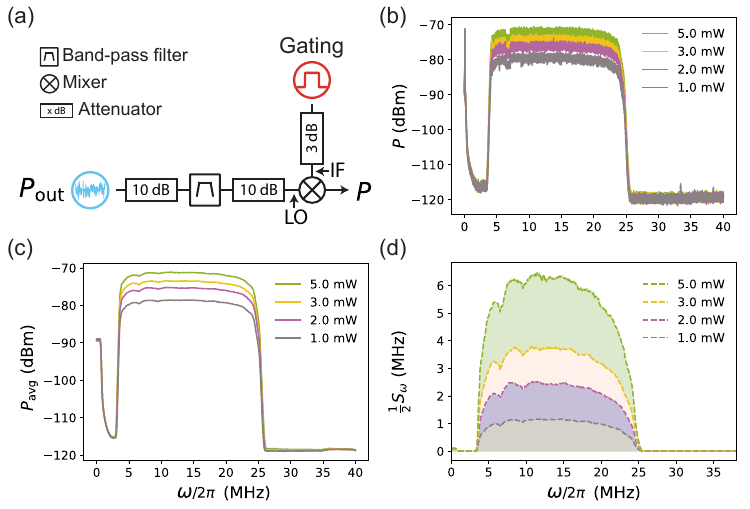}
\caption{\label{suppfig:Noise}
Noise generation, characterization, and impact. (a) Schematic of the electronic setup for generating classical noise with programmable spectral and temporal profiles. (b) Measured noise power spectrum (averaged over 100 traces) showing the filtered band-limited profile. The dip near 6\,MHz is an artifact of the measurement apparatus. (c) Processed spectrum obtained by applying a robust moving-window average (450-point window, selecting the 30th point from the maximum) to the data in (b). (d) Calculated power spectral density of frequency noise affecting the qubit (derived from Eq.\,(\ref{eq:somega})) for different injected noise powers $P_{\mathrm{out}}$. Dashed lines represent theoretical predictions based on the characterized noise injection chain.
}
\end{figure}

In this work, we utilize an AWG as a controllable source of classical noise, with its output power denoted as $P_{\mathrm{out}}$. The experimental setup is illustrated schematically in Fig.\,\ref{suppfig:Noise}(a). The AWG inherently generates broadband noise with an intrinsic bandwidth of $\pm 90\,\mathrm{MHz}$.

To tailor the spectral profile of this noise, the raw AWG output is first directed through a band-pass filter with a nominal passband of $5$–$20\,\mathrm{MHz}$. This filtering stage confines the effective noise bandwidth to this designated frequency range, thereby suppressing out-of-band spectral components. Subsequently, to achieve precise temporal modulation of the noise, the filtered signal is combined with a gating waveform via an RF mixer. The gating waveform, which defines the envelope of the noise pulse in time, is supplied by a separate ``Gating" control line (see Fig.\,\ref{suppfig:Noise}(a)). The mixer employed has a local oscillator (LO) frequency range of $0.01$–$6\,\mathrm{GHz}$ and supports an intermediate frequency (IF) bandwidth from DC to $1\,\mathrm{GHz}$, enabling flexible waveform synthesis.

The spectro-temporal characteristics of the modulated noise signal are verified using a spectrum analyzer. For spectral characterization, the gating line is held at a constant DC voltage of $0.56\,\mathrm{V}$, resulting in a continuous noise output. The measured power spectrum $P$, acquired by averaging over 100 independent traces, is presented in Fig.\,\ref{suppfig:Noise}(b) for the frequency range $0$–$40\,\mathrm{MHz}$. The resolution bandwidth (RBW) of the spectrum analyzer was set to $1\,\mathrm{kHz}$ for these measurements. A distinctive dip observed near $6\,\mathrm{MHz}$ is attributed to an instrumental artifact of the spectrum analyzer and is not a feature of the generated noise.

To extract a smoothed spectral trend, the data from Fig.\,\ref{suppfig:Noise}(b) are further processed via a moving-average procedure. For each frequency point, a local window encompassing 450 adjacent power ($P$) values is considered. Within each window, the $30^{\mathrm{th}}$ highest power value (counting from the maximum) is selected as a robust estimator, denoted as $P_{\mathrm{avg}}$. The resultant averaged spectrum, with $P_{\mathrm{avg}}$ expressed in $\mathrm{dBm}$, is displayed in Fig.\,\ref{suppfig:Noise}(c).

The voltage noise power spectral density (PSD) experienced by the qubit at its operating frequency can be derived from the measured output power $P_{\mathrm{avg}}$ using the following relation:
\begin{equation}
S_{\omega} = C \cdot \frac{10^{(P_{\rm avg} - 30)/10}}{\mathrm{RBW}} \cdot R \cdot \left( 2\pi \cdot \frac{\partial f}{\partial V} \right)^2 .
\label{eq:somega}
\end{equation}
Here, $\mathrm{RBW} = 10^3 \, \mathrm{Hz}$ is the resolution bandwidth of the spectrum analyzer, $R = 50 \, \Omega$ is the characteristic impedance of the line, and $\frac{\partial f}{\partial V}$ represents the sensitivity of the qubit transition frequency to an applied voltage on its control line. The dimensionless prefactor $C$ aggregates the total attenuation (or gain) between the point where $P_{\mathrm{avg}}$ is measured and the point where the qubit frequency sensitivity $\frac{\partial f}{\partial V}$ is characterized.

Prior to entering the cryogenic system, the modulated noise pulse is subjected to a $20\,\mathrm{dB}$ attenuator. It is then combined with the qubit's DC Z-control bias via a power splitter, which introduces an additional insertion loss of $6.3\,\mathrm{dB}$. The qubit's frequency sensitivity $\frac{\partial f}{\partial V}$ was characterized independently via spectroscopy through the Z-control line (with a $10\,\mathrm{dB}$ attenuator in place) and was found to be approximately $1.9 \times 10^9 \, \mathrm{Hz/V}$. Using this value and accounting for all known losses in the signal path, the aggregate constant $C$ is calculated to be $10^{-1.63}$.

Substituting the measured $P_{\mathrm{avg}}$ spectrum, along with the parameters $\mathrm{RBW}$, $R$, $\frac{\partial f}{\partial V}$, and $C$ into Eq.\,(\ref{eq:somega}), yields the theoretically predicted frequency noise PSD, $S_{\omega}$. The corresponding decoherence rate induced by this classical noise is given by $\frac{1}{2}S_{\omega}$. The frequency dependence of this theoretically predicted dissipation rate is plotted with dashed lines in Fig.\,\ref{suppfig:Noise}(d) for various injected noise power settings $P_{\mathrm{out}}$.

\subsection{Calibration of the inter-qubit coupling strength}

The effective coupling strength $g$ between the two transmon qubits is calibrated through a time-domain measurement of coherent swap oscillations under resonant conditions. The calibration procedure is executed as follows. First, qubit $\mathrm{Q}_0$ is prepared in its first excited state $\ket{1}$ via a resonant $\pi$-pulse, while the neighboring qubit $\mathrm{Q}_1$ is initialized in the ground state $\ket{0}$. Subsequently, both qubits are tuned to the same frequency bringing them into mutual resonance.

\begin{figure}[tbhp]
\centering
\includegraphics[scale=1]{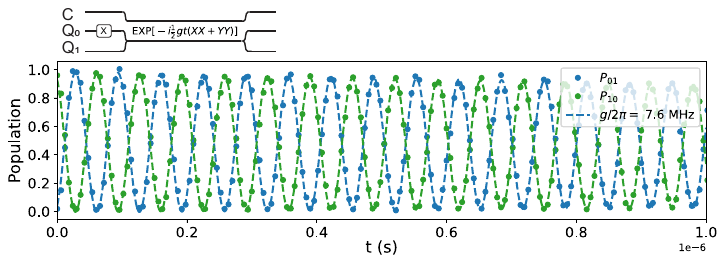}
\caption{\label{suppfig:calig} 
Calibration of the qubit--qubit coupling strength. Qubit $\mathrm{Q}_0$ is initialized in $\ket{1}$ and $\mathrm{Q}_1$ in $\ket{0}$. With both qubits tuned to resonance, the excitation coherently swaps between them as the interaction time $t$ is varied. The joint two-qubit measurement yields oscillating populations of the $\ket{10}$ and $\ket{01}$ states (data points). A damped sinusoidal fit (solid line) to either population yields the swap frequency $f_{\mathrm{swap}}$, from which the coupling strength $2g/2\pi = f_{\mathrm{swap}}$ is determined.
}
\end{figure}

Under this resonant condition, the system is governed by the exchange Hamiltonian $H = g (\sigma_+^{(0)} \sigma_-^{(1)} + \sigma_-^{(0)} \sigma_+^{(1)})$, where $\sigma_\pm^{(j)}$ are the raising and lowering operators for qubit $j$, and $g$ denotes the effective coherent coupling strength between the two qubits. This Hamiltonian drives coherent oscillations between the two-qubit states $\ket{10}$ and $\ket{01}$, analogous to Rabi oscillations in a two-level system. As a function of the interaction time $t$, the excited-state population of either qubit exhibits sinusoidal swapping behavior. Experimentally, we vary the duration $t$ of the resonant interaction and perform joint two-qubit readout to directly measure the populations of the $\ket{10}$ and $\ket{01}$ states (or equivalently, the probability $P_{10}(t)$ for $\mathrm{Q}_0$ being excited while $\mathrm{Q}_1$ is in the ground state). The recorded data, presented in Fig.\,\ref{suppfig:calig}, display clear swap oscillations.

The oscillation frequency is directly related to the coupling strength via $f_{\mathrm{swap}} = 2g/2\pi$. By fitting the time-domain population data to a damped sinusoidal function of the form $P_{01}(t) = A e^{-t/\tau} \cos(2\pi f_{\mathrm{swap}} t + \phi) + B$, we extract $f_{\mathrm{swap}}$ and thus obtain $2g/2\pi=f_{\rm swap}$. This method provides a direct, time-domain measurement of $g$ that is independent of specific models for the coupler or detailed circuit parameters.


\subsection{Numerical simulation of gate error under injected flux noise}

\begin{figure}[tbhp]
\centering
\includegraphics[scale=1]{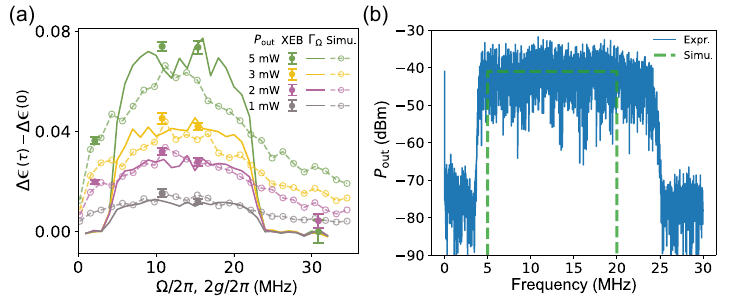}
\caption{\label{suppfig:simu_err}
Comparison of experimental and simulated gate errors. (a) The change in gate error $\Delta\epsilon$ for a fixed interaction duration $\tau = 48\,\mathrm{ns}$ is plotted as a function of the Rabi frequency $\Omega$ and the coupling strength $g$ for different injected noise powers. Solid lines indicate the theoretically predicted $\Delta\epsilon$ derived from the experimentally measured dephasing rate $\Gamma_{\Omega}$. Solid dots correspond to gate errors measured directly via XEB experiments. Hollow circles represent the results of our numerical simulations. (b) Spectral characteristics of the injected noise. The blue trace shows the frequency response of the 5–20\,MHz bandpass filter used experimentally. The dashed green trace shows the idealized, flat noise spectrum implemented in the numerical simulations.
}
\end{figure}

We performed comprehensive numerical simulations to model the expected gate error under the influence of the characterized flux noise. The simulated cross-entropy-benchmarked (XEB) gate error, against the experimental XEB protocol, is plotted as hollow circles in Fig.\,\ref{suppfig:simu_err}(a).

Within the central region of the passband (approximately 12.5\,MHz), the simulation results show good agreement with the experimentally measured errors (solid dots). This consistency validates our noise characterization and modeling framework for the dominant noise components near the band center. Outside this spectral region, however, a discrepancy emerges between the simulation and experiment. 
It is because the noise frequencies in this range are largely non-resonant with the relevant system dynamics (primarily the $2g$ frequency associated with the qubit-qubit interaction), the decoherence process becomes more complex. This non-resonant dissipative mechanism warrants detailed theoretical investigation, which we identify as an important direction for future work and was not explored in depth in the present study.

The spectral profile of the noise used in the simulation is illustrated in Fig.\,\ref{suppfig:simu_err}(b). The blue curve represents the transfer function of the actual 5–20\,MHz bandpass filter employed in the experiment. The dashed green line depicts the idealized, flat frequency-domain response of the additive noise that was implemented in our simulations, serving as the input for the theoretical error calculations.

\end{document}